\documentclass[3p,times,procedia]{elsarticle}
\usepackage{ecrc}

\volume{00}

\firstpage{1}

\journalname{Transportation Research Procedia}

\runauth{J. Raimbault}

\jid{aaspro}

\usepackage{amssymb}

\biboptions{authoryear}

\usepackage[figuresright]{rotating}

\usepackage{bbm}

\begin{document}

\begin{frontmatter}



\dochead{19th EURO Working Group on Transportation Meeting, EWGT2016, 5-7 September 2016, Istanbul, Turkey}

\title{Investigating the Empirical Existence of Static User Equilibrium}


\author[a,b]{Juste Raimbault\corref{cor1}}

\address[a]{UMR CNRS 8504 G{\'e}ographie-cit{\'e}s, 13 rue du Four, 75006 Paris, France}
\address[b]{UMR-T IFSTTAR 9403 LVMT, Cit{\'e} Descartes, 77455 Champs-sur-Marne, France}

\begin{abstract}
The Static User Equilibrium is a powerful framework for the theoretical study of traffic. Despite the restricting assumption of stationary flows that intuitively limit its application to real traffic systems, many operational models implementing it are still used without an empirical validation of the existence of the equilibrium. We investigate its existence on a traffic dataset of three months for the region of Paris, FR. The implementation of an application for interactive spatio-temporal data exploration allows to hypothesize a high spatial and temporal heterogeneity, and to guide further quantitative work. The assumption of locally stationary flows is invalidated in a first approximation by empirical results, as shown by a strong spatial and temporal variability in shortest paths and in network topological measures such as betweenness centrality. Furthermore, the behavior of spatial autocorrelation index of congestion patterns at different spatial ranges suggest a chaotic evolution at the local scale, especially during peak hours. We finally discuss the implications of these empirical findings and describe further possible developments based on the estimation of Lyapunov dynamical stability of traffic flows.
\end{abstract}

\cortext[cor1]{Corresponding author. Tel.: +33140464000.
}
\end{frontmatter}

\email{juste.raimbault@polytechnique.edu}




\section{Introduction}
\label{introduction}

Traffic Modeling has been extensively studied since seminal work by~\cite{wardrop1952road} : economical and technical elements at stake justify the need for a fine understanding of mechanisms ruling traffic flows at different scales. Many approaches with different purposes coexist today, of which we can cite dynamical micro-simulation models, generally opposed to equilibrium-based techniques. Whereas the validity of micro-based models has been largely discussed and their application often questioned, the literature is relatively poor on empirical studies assessing the stationary equilibrium assumption in the Static User Equilibrium (SUE) framework. Various more realistic developments have been documented in the literature, such as Dynamic Stochastic User Equilibrium (DSUE) (see e.g. a description by~\cite{han2003dynamic}). An intermediate between static and stochastic frameworks is the Restricted Stochastic User Equilibrium, for which route choice sets are constrained to be realistic (\cite{rasmussen2015stochastic}). Extensions that incorporate user behavior with choice models have more recently been proposed, such as~\cite{zhang2013dynamic} taking into account both the influence of road pricing and congestion on user choice with a Probit model. Relaxations of other restricting assumptions such as pure user utility maximization have been also introduced, such as the Boundedly Rational User Equilibrium described by~\cite{mahmassani1987boundedly}. In this framework, user have a range of satisfying utilities and equilibrium is achieved when all users are satisfied. It produces more complex features such as the existence of multiple equilibria, and allows to account for specific stylized facts such as irreversible network change as developed by~\cite{guo2011bounded}. Other models for traffic assignment, inspired from other fields have also recently been proposed : in~\cite{puzis2013augmented}, an extended definition of betweenness centrality combining linearly free-flow betweenness with travel-time weighted betweenness yield a high correlation with effective traffic flows, acting thus as a traffic assignment model. It provides direct practical applications such as the optimization of traffic monitors spatial distribution.

Despite all these developments, some studies and real-world applications still rely on Static User Equilibrium. Parisian region e.g. uses a static model (MODUS) for traffic management and planning purposes. \cite{leurent2014user} introduce a static model of traffic flow including parking cruising and parking lot choice: it is legitimate to ask, specifically at such small scales, if the stationary distribution of flows is a reality. An example of empirical investigation of classical assumptions is given in~\cite{zhu2010people}, in which revealed route choices are studied. Their conclusions question ``Wardrop’s first principle'' implying that users choose among a well-known set of alternatives. In the same spirit, we investigate the possible existence of the equilibrium in practice. More precisely, SUE assumes a stationary distribution of flows over the whole network. This assumption stays valid in the case of local stationarity, as soon as time scale for parameter evolution is considerably greater than typical time scales for travel. The second case which is more plausible and furthermore compatible with dynamical theoretical frameworks, is here tested empirically.

The rest of the paper is organized as follows : data collection procedure and dataset are described ; we present then an interactive application for the interactive exploration of the dataset aimed to give intuitive insights into data patterns ; we present then results of various quantitative analyses that give convergent evidence for the non-stationarity of traffic flows ; we finally discuss implications of these results and possible developments.

\section{Data collection}

\subsection{Dataset Construction}

We propose to work on the case study of Parisian Metropolitan Region. An open dataset was constructed for highway links within the region, collecting public real-time open data for travel times (available at www.sytadin.fr). As stated by~\cite{bouteiller2013open}, the availability of open datasets for transportation is far to be the rule, and we contribute thus to a data opening by the construction of our dataset. Our data collection procedure consists in the following simple steps, executed each two minutes by a \texttt{python} script :
\begin{itemize}
\item fetch raw webpage giving traffic information
\item parse html code to retrieve traffic links id and their corresponding travel time
\item insert all links in a \texttt{sqlite} database with the current timestamp.
\end{itemize}
The automatized data collection script continues to enrich the database as time passes, allowing future extensions of this work on a larger dataset and a potential reuse by scientists or planners. The latest version of the dataset is available online (sqlite format) under a Creative Commons License\footnote{at \texttt{http://37.187.242.99/files/public/sytadin{\_}latest.sqlite3}}.

\subsection{Data Summary}

A time granularity of 2 minutes was obtained for a three months period (February 2016 to April 2016 included). Spatial granularity is in average 10km, as travel times are provided for major links. The dataset contains 101 links. Raw data we use is effective travel time, from which we can construct travel speed and relative travel speed, defined as the ratio between optimal travel time (travel time without congestion, taken as minimal travel times on all time steps) and effective travel time. Congestion is constructed by inversion of a simple BPR function with exponent 1, i.e. we take $c_i = 1 - \frac{t_{i,min}}{t_i}$ with $t_i$ travel time in link $i$ and $t_{i,min}$ minimal travel time.

\section{Methods and Results}

\subsection{Visualization of spatio-temporal congestion patterns}

\begin{figure}
\vspace{1cm}
\centering
\includegraphics[width=\textwidth]{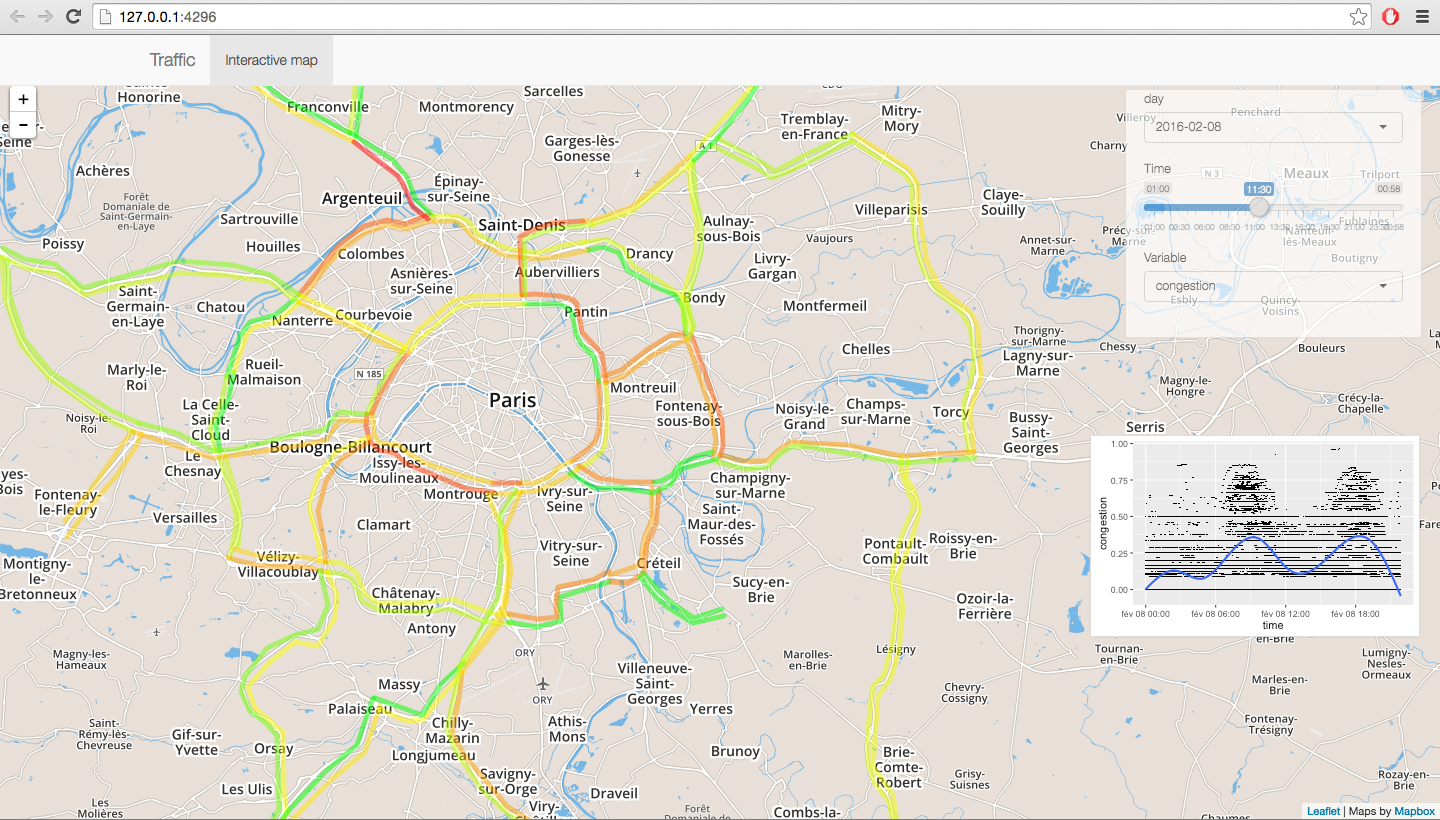}
\caption{Capture of the web-application to explore spatio-temporal traffic data for Parisian region. It is possible to select date and time (precision of 15min on one month, reduced from initial dataset for performance purposes). A plot summarizes congestion patterns on the current day.}
\label{fig:fig-1}
\end{figure}

As our approach is fully empirical, a good knowledge of existing patterns for traffic variables, and in particular of their spatio-temporal variations, is essential to guide any quantitative analysis. Taking inspiration from an empirical model validation literature, more precisely Pattern-oriented Modeling techniques introduced by~\cite{grimm2005pattern}, we are interested in macroscopic patterns at given temporal and spatial scales: the same way stylized facts are in that approach extracted from a system before trying to model it, we need to explore interactively data in space and time to find relevant patterns and associated scales. We implemented therefore an interactive web-application for data exploration using \texttt{R} packages \texttt{shiny} and \texttt{leaflet}\footnote{source code for the application and analyses is available on project open repository at\\
\texttt{https://github.com/JusteRaimbault/TransportationEquilibrium}}.
It allows dynamical visualization of congestion among the whole network or in a particular area when zoomed in. The application is accessible online at \texttt{http://shiny.parisgeo.cnrs.fr/transportation}. A screenshot of the interface is presented in Figure~\ref{fig:fig-1}. Main conclusion from interactive data exploration is that strong spatial and temporal heterogeneity is the rule. The temporal pattern recurring most often, peak and off-peak hours is on a non-negligible proportion of days perturbed. In a first approximation, non-peak hours may be approximated by a local stationary distribution of flows, whereas peaks are too narrow to allow the validation of the equilibrium assumption. Spatially we can observe that no spatial pattern is clearly emerging. It means that in case of a validity of static user equilibrium, meta-parameters ruling its establishment must vary at time scales smaller than one day. We argue that traffic system must in contrary be far-from-equilibrium, especially during peak hours when critical phase transitions occur at the origin of traffic jams.

\subsection{Spatio-temporal Variability of Travel Path}

\begin{figure}
\centering
\vspace{1.5cm}
\includegraphics[width=0.47\textwidth]{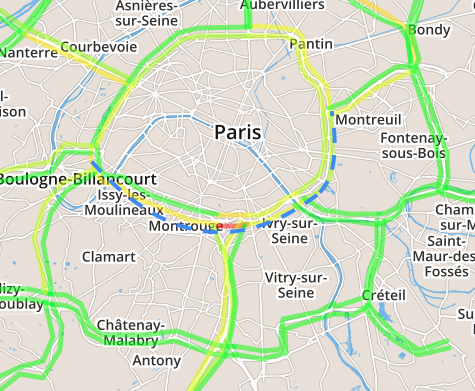}\hfill
\includegraphics[width=0.47\textwidth]{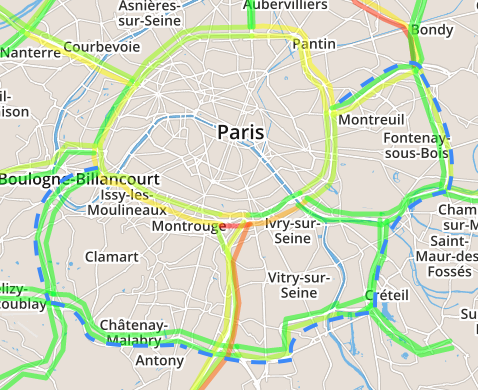}
\caption{Spatial variability of travel-time shortest path (shortest path trajectory in dotted blue). In an interval of only 10 minutes, between 11/02/2016 00:06 (left) and 11/02/2016 00:16 (right), the shortest path between \emph{Porte d'Auteuil} (West) and \emph{Porte de Bagnolet} (East), increases in effective distance of $\simeq 37$km (with an increase in travel time of only 6min), due to a strong disruption on the ring of Paris.} 
\label{fig:fig-2}
\end{figure}

Following interactive exploration of data, we propose to quantify the spatial variability of congestion patterns to validate or invalidate the intuition that if equilibrium does exist in time, it is strongly dependent on space and localized. The variability in time and space of travel-time shortest paths is a first way to investigate flow stationarities from a game-theoretic point of view. Indeed, the static User Equilibrium is the stationary distribution of flows under which no user can improve its travel time by changing its route. A strong spatial variability of shortest paths at short time scales is thus evidence of non-stationarity, since a similar user will take a few time after a totally different route and not contribute to the same flow as a previous user. Such a variability is indeed observed on a non-negligible number of paths on each day of the dataset. We show in Figure~\ref{fig:fig-2} an example of extreme spatial variation of shortest path for a particular Origin-Destination pair.

The systematic exploration of travel time variability across the whole dataset, and associated travel distance, confirms, as described in Figure 3, that travel time absolute variability has often high values of its maximum across OD pairs, up to 25 minutes with a temporal local mean around 10min. Corresponding spatial variability produces detours up to 35km.

\begin{figure}[t]\vspace*{4pt}
\centering
\centerline{\includegraphics[width=0.8\textwidth]{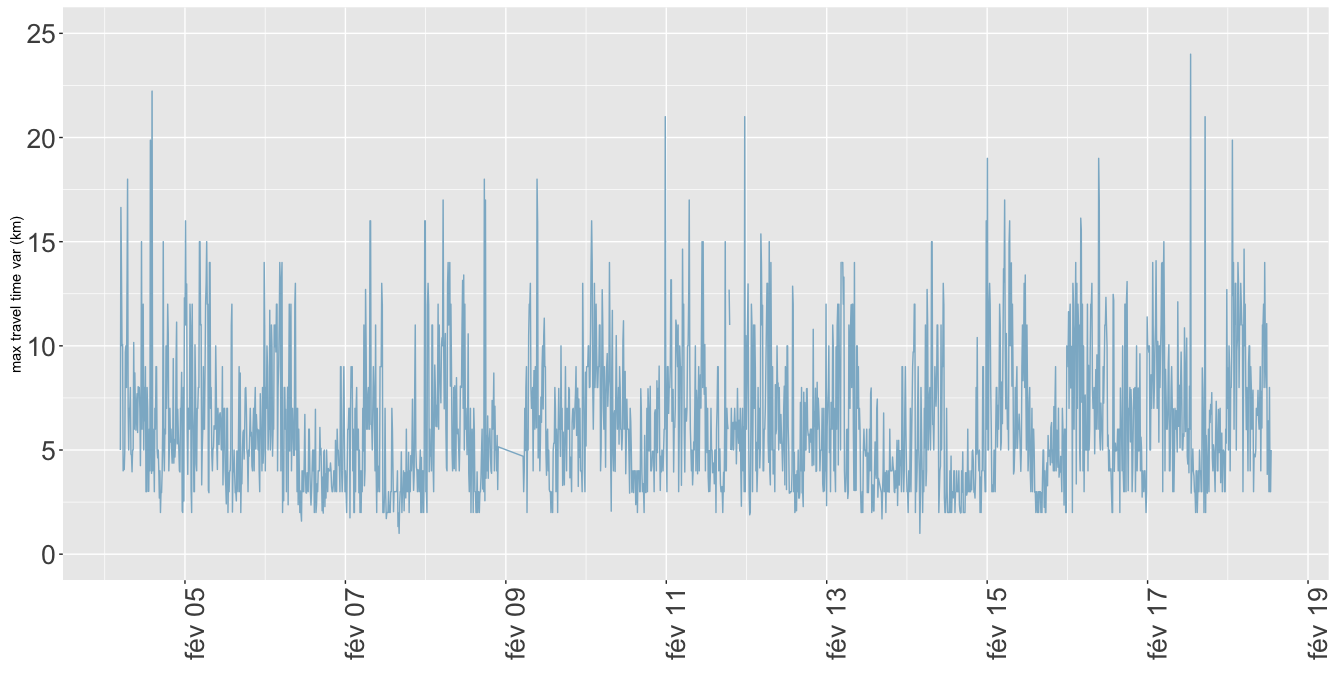}}
\centerline{\includegraphics[width=0.8\textwidth]{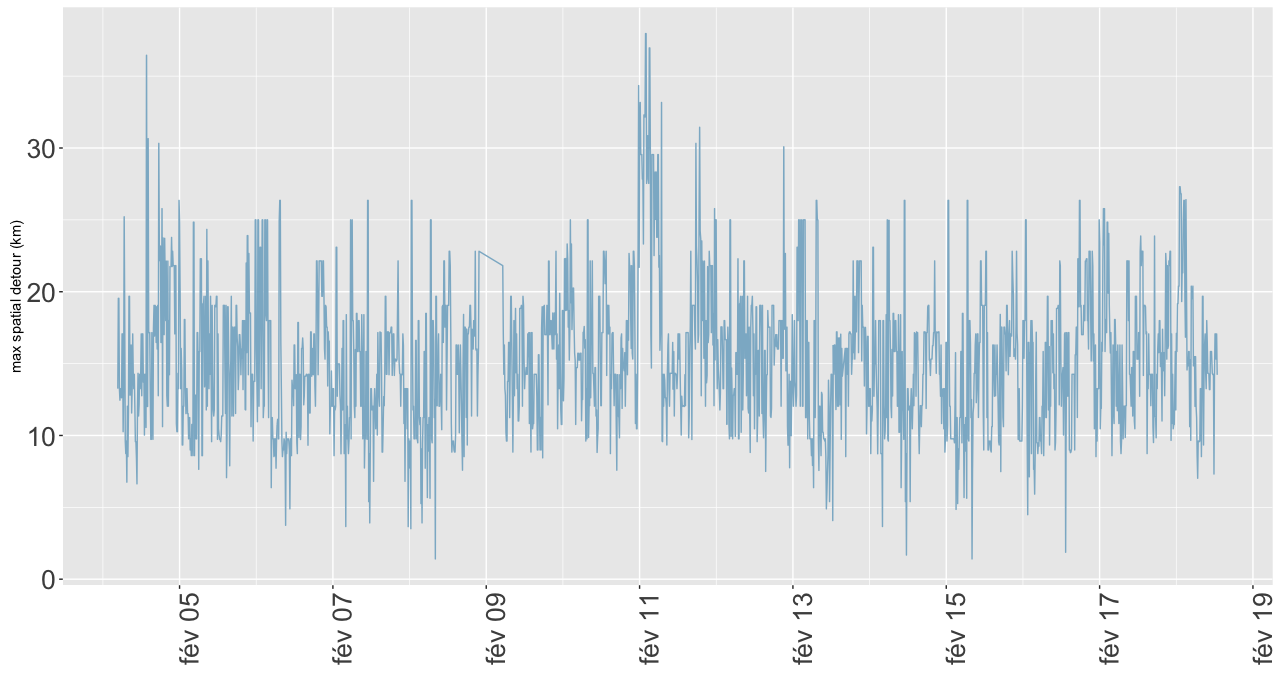}}
\caption{Travel time (top) in min and corresponding travel distance (bottom) maximal variability on a two weeks sample. We plot the maximal on all OD pairs of the absolute variability between two consecutive time steps. Peak hours imply a high time travel variability up to 25 minutes and a path length variability up to 35km.}
\label{fig:fig-3}
\end{figure}

\subsection{Stability of Network measures}

The variability of potential trajectories observed in the previous section can be confirmed by studying the variability of network properties. In particular, network topological measures capture global patterns of a transportation network. Centrality and node connectivity measures are classical indicators in transportation network description as recalled in~\cite{bavoux2005geographie}. The transportation literature has developed elaborated and operational network measures, such as network robustness measures to identify critical links and measure overall network resilience to disruptions (an example among many is the Network Trip Robustness index introduced in~\cite{sullivan2010identifying}).

More precisely, we study the betweenness centrality of the transportation network, defined for a node as the number of shortest paths going through the node, i.e. by the equation

\begin{equation}
b_i = \frac{1}{N(N-1)}\cdot \sum_{o\neq d \in V}\mathbbm{1}_{i\in p(o\rightarrow d)}
\end{equation}

where $V$ is the set of network vertices of size $N$, and $p(o\rightarrow d)$ is the set of nodes on the shortest path between vertices o and d (the shortest path being computed with effective travel times). This index is more relevant to our purpose than other measures of centrality such as closeness centrality that does not include potential congestion as betweenness centrality does.

\begin{figure}
\includegraphics[width=\textwidth]{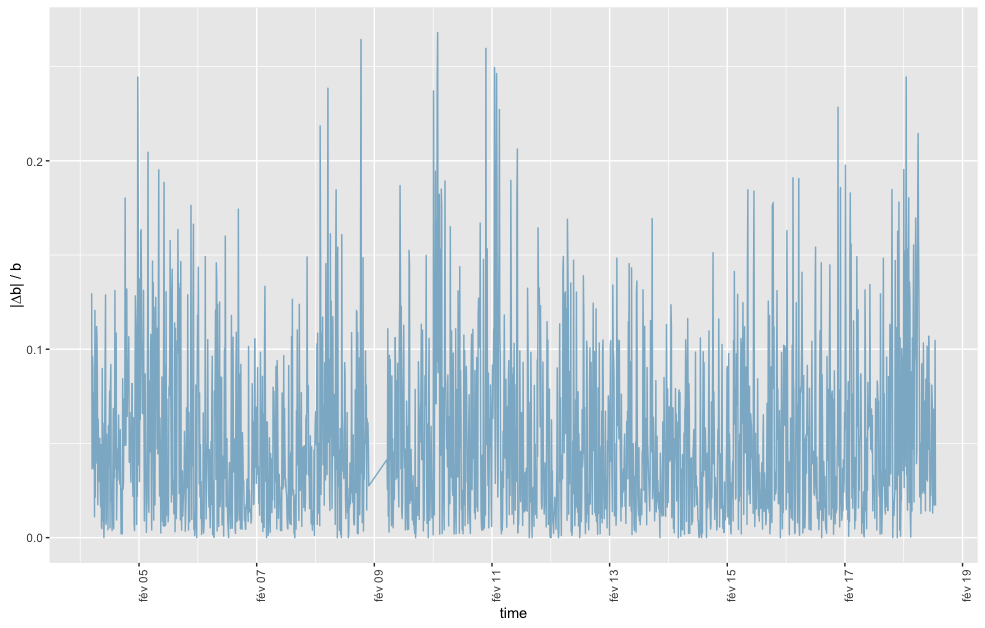}
\caption{Temporal stability of maximal betweenness centrality. We plot in time the normalized derivative of maximal betweenness centrality, that expresses its relative variations at each time step. The maximal value up to 25\% correspond to very strong network disruption on the concerned link, as it means that at least this proportion of travelers assumed to take this link in previous conditions should take a totally different path.}
\label{fig:fig-4}
\end{figure}

We show in Figure 4 the relative absolute variation of maximal betweenness centrality for the same time window than previous empirical indicators. More precisely we plot the value of

\begin{equation}
\Delta b(t) = \frac{\left|\max_i (b_i(t + \Delta t)) - \max_i (b_i(t))\right|}{\max_i (b_i(t))}
\end{equation}

where $\Delta t$ is the time step of the dataset (the smallest time window on which we can capture variability). This absolute relative variation has a direct meaning : a variation of 20\% (which is attained a significant number of times as shown in Fig.~\ref{fig:fig-4}) means that in case of a negative variation, at least this proportion of potential travels have changed route and the local potential congestion has decrease of the same proportion. In the case of a positive variation, a single node has captured at least 20\% of travels. Under the assumption (that we do not try to verify in this work and assume to be also not verified as shown by~\cite{zhu2010people}, but that we use as a tool to give an idea of the concrete meaning of betweenness variability) that users rationally take the shortest path and assuming that a majority of travels are realized such a variation in centrality imply a similar variation in effective flows, leading to the conclusion that they can not be stationary in time (at least at a scale larger than $\Delta t$) nor in space.

\subsection{Spatial heterogeneity of equilibrium}

To obtain a different insight into spatial variability of congestion patterns, we propose to use an index of spatial autocorrelation, the Moran index (defined e.g. in~\cite{tsai2005quantifying}). More generally used in spatial analysis with diverse applications from the study of urban form to the quantification of segregation, it can be applied to any spatial variable. It allows to establish neighborhood relations and unveils spatial local consistence of an equilibrium if applied on localized traffic variable. At a given point in space, local autocorrelation for variable c is computed by

\begin{equation}
\rho_i = \frac{1}{K}\cdot \sum_{i\neq j}{w_{ij}\cdot (c_i - \bar{c})(c_j - \bar{c})}
\end{equation}

where $K$ is a normalization constant equal to the sum of spatial weights times variable variance and $\bar{c}$ is variable mean. In our case, we take spatial weights of the form $w_{ij} = \exp{\left(\frac{-d_{ij}}{d_0}\right)}$ with $d_0$ typical decay distance and compute the autocorrelation of link congestion localized at link center. We capture therefore spatial correlations within a radius of same order than decay distance around the point $i$. The mean on all points yields spatial autocorrelation index $I$. A stationarity in flows should yield some temporal stability of the index.

\begin{figure}
\includegraphics[width=\textwidth,height=0.6\textheight]{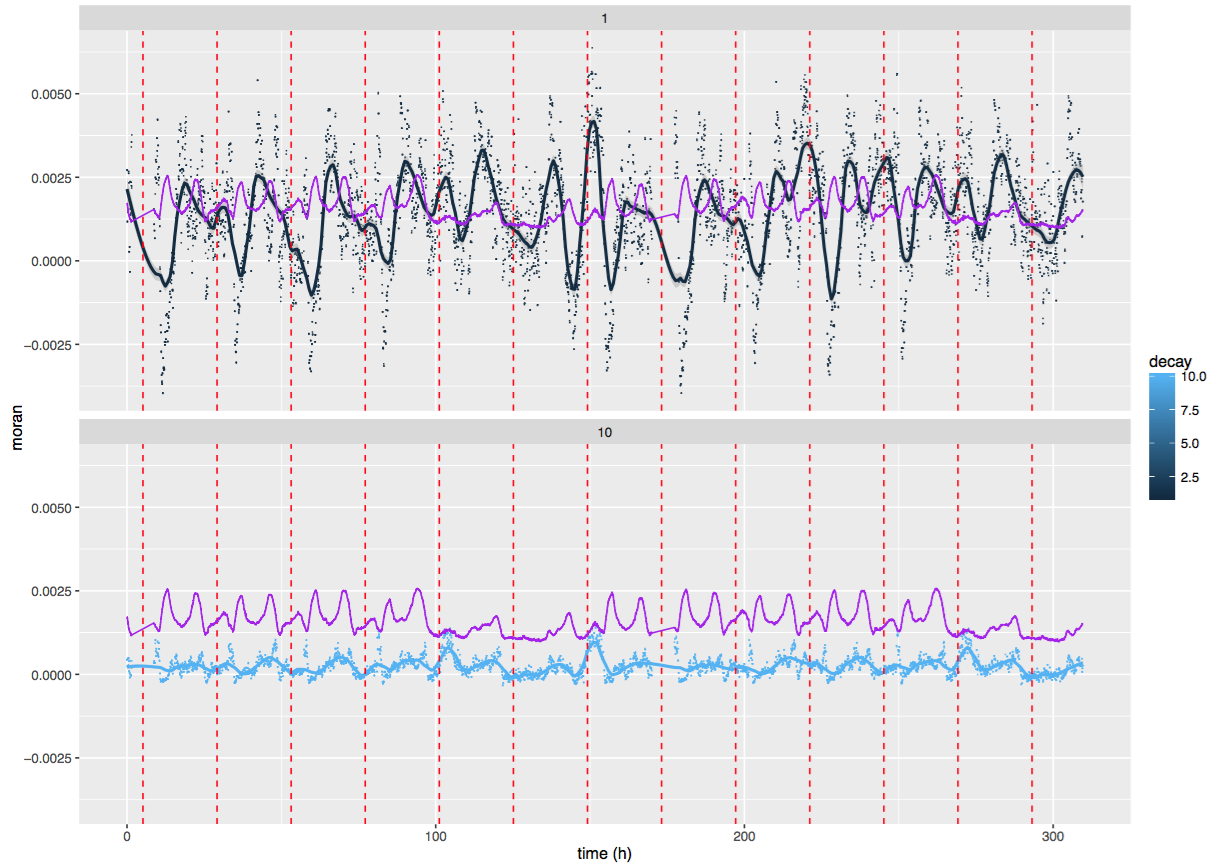}
\caption{Spatial auto-correlations for relative travel speed on two weeks. We plot for varying value of decay parameter (1,10km) values of auto-correlation index in time. Intermediate values of decay parameter yield a rather continuous deformation between the two curves. Points are smoothed with a 2h span to ease reading. Vertical dotted lines correspond to midnight each day. Purple curve is relative speed fitted at scale to have a correspondence between auto-correlation variations and peak hours.}
\label{fig:fig-5}
\end{figure}

Figure~\ref{fig:fig-5} presents temporal evolution of spatial autocorrelation for congestion. As expected, we have a strong decrease of autocorrelation with distance decay parameter, for both amplitude and temporal average. The high temporal variability implies short time scales for potential stationarity windows. When comparing with congestion (fitted to plot scale for readability) for 1km decay, we observe that high correlations coincide with off-peak hours, whereas peaks involve vanishing correlations. Our interpretation, combined with the observed variability of spatial patterns, is that peak hours correspond to chaotic behaviour of the system, as jams can emerge in any link: correlation thus vanishes as feasible phase space for a chaotic dynamical system is filled by trajectories in an uniform way what is equivalent to apparently independent random relative speeds.

\section{Discussion}

\subsection{Theoretical and practical implications of empirical conclusions}

We argue that the theoretical implications of our empirical findings do not imply in a total discarding of the Static User Equilibrium framework, but unveil more a need of stronger connections between theoretical literature and empirical studies. If each newly introduced theoretical framework is generally tested on one on more case study, there are no systematic comparisons of each on large and different datasets and on various objectives (prediction of traffic, reproduction of stylized facts, etc.) as systematic reviews are the rule in therapeutic evaluation for example. This imply however broader data and model sharing practices than the current ones. The precise knowledge of application potentialities for a given framework may induce unexpected developments such as its integration into larger models. The example of Land-use and Transportation Interaction studies (LUTI models) is a good illustration of how the SUE can still be used for larger purpose than transportation modeling. \cite{kryvobokov2013comparison} describe two LUTI models, one of which includes two equilibria for four-step transportation model and for land-use evolution (households and firms relocation), the other being more dynamical. The conclusion is that each model has its own advantages regarding the pursued objective, and that the static model can be used for long time policy purposes, whereas the dynamic model provide more precise information at smaller time scale. In the first case, a more complicated transportation module would have been complicated to include, what is an advantage of the static user equilibrium.

Concerning practical applications, it seems natural that static models should not be used for traffic forecast and management at small time scales (week or day) and efforts should be made to implement more realistic models. However the use of models by the planning and engineering community is not necessarily directly related to academic concerns and state-of-the-art. For the particular case of France and mobility models, \cite{commenges2013invention} showed that engineers had gone to the point of constructing inexistent problems and implementing corresponding models that they had imported from a totally different geographical context (planning in the United States). The use of one framework or type of model has historical reasons that may be difficult to overcome.

\subsection{Towards explanative interpretations of non-stationarity}

An assumption we formulate regarding the origin of non-stationarity of network flows, in view of data exploration and quantitative analysis of the database, is that the network is at least half of the time highly congested and in a critical state. The off-peak hours are the larger potential time windows of spatial and temporal stationarity, but consist in less than half of the time. As already interpreted through the behavior of autocorrelation indicator, a chaotic behavior may be at the origin of such variability in the congested hours. The same way a supercritical fluid may condense under the smallest external perturbation, the state of the link may qualitatively change with a small incident, producing a network disruption that may propagate and even amplify. The direct effect of traffic events (notified incidents or accidents) can not be studied without external data, and it could be interesting to enrich the database in that direction. It would allow establishing the proportion of disruptions that do appear to have a direct effect and quantify a level of criticality of network congestion in time, or to investigate more precise effects such as the consequences of an incident on traffic of the opposite lane.

\subsection{Possible developments}

Further work may be planned towards a more refined assessment of temporal stability on a region of the network, i.e. the quantitative investigation of consideration of peak stationarity given above. To do so we propose to compute numerically Liapounov stability of the dynamical system ruling traffic flows using numerical algorithms such as described by~\cite{goldhirsch1987stability}. The value of Liapounov exponents provides the time scale by which the unstable system runs out of equilibrium. Its comparison with peak duration and average travel time, across different spatial regions and scales should provide more information on the possible validity of the local stationarity assumption. This technique has already been introduced at an other scale in transportation studies, as e.g.~\cite{tordeux2016jam} that study the stability of speed regulation models at the microscopic scale to avoid traffic jams.

Other research directions may consist in the test of other assumptions of static user equilibrium (as the rational shortest path choice, which would be however difficult to test on such an aggregated dataset, implying the use of simulation models calibrated and cross-validated on the dataset to compare assumptions, without necessarily a direct clear validation or invalidation of the assumption), or the empirical computation of parameters in stochastic or dynamical user equilibrium frameworks.

\section{Conclusion}

We have described an empirical study aimed at a simple but from our point of view necessary investigation of the existence of the static user equilibrium, more precisely of its stationarity in space and time on a metropolitan highway network. We constructed by data collection a traffic congestion dataset for the highway network of Greater Paris on 3 months with two minutes temporal granularity. The interactive exploration of the dataset with a web application allowing spatio-temporal data visualization helped to guide quantitative studies. Spatio-temporal variability of shortest paths and of network topology, in particular betweenness centrality, revealed that stationarity assumptions do not hold in general, what was confirmed by the study of spatial autocorrelation of network congestion. We suggest that our findings highlight a general need of higher connections between theoretical and empirical studies, as our work can discard misunderstandings on the theoretical static user equilibrium framework and guide the choice of potential applications.



\bibliography{biblio}

\newcommand{\noopsort}[1]{} \newcommand{\printfirst}[2]{#1}
  \newcommand{\singleletter}[1]{#1} \newcommand{\switchargs}[2]{#2#1}
\begin{thebibliography}{18}
\expandafter\ifx\csname natexlab\endcsname\relax\def\natexlab#1{#1}\fi
\providecommand{\url}[1]{\texttt{#1}}
\providecommand{\href}[2]{#2}
\providecommand{\path}[1]{#1}
\providecommand{\DOIprefix}{doi:}
\providecommand{\ArXivprefix}{arXiv:}
\providecommand{\URLprefix}{URL: }
\providecommand{\Pubmedprefix}{pmid:}
\providecommand{\doi}[1]{\href{http://dx.doi.org/#1}{\path{#1}}}
\providecommand{\Pubmed}[1]{\href{pmid:#1}{\path{#1}}}
\providecommand{\bibinfo}[2]{#2}
\ifx\xfnm\relax \def\xfnm[#1]{\unskip,\space#1}\fi
\bibitem[{Bavoux et~al.(2005)Bavoux, Beaucire, Chapelon and
  Zembri}]{bavoux2005geographie}
\bibinfo{author}{Bavoux, J.J.}, \bibinfo{author}{Beaucire, F.},
  \bibinfo{author}{Chapelon, L.}, \bibinfo{author}{Zembri, P.},
  \bibinfo{year}{2005}.
\newblock \bibinfo{title}{G{\'e}ographie des transports}.
\newblock \bibinfo{publisher}{Paris}.
\bibitem[{Bouteiller and Berjoan(2013)}]{bouteiller2013open}
\bibinfo{author}{Bouteiller, C.}, \bibinfo{author}{Berjoan, S.},
  \bibinfo{year}{2013}.
\newblock \bibinfo{title}{Open data en transport urbain: quelles sont les
  donn{\'e}es mises {\`a} disposition? quelles sont les strat{\'e}gies des
  autorit{\'e}s organisatrices?} .
\bibitem[{Commenges(2013)}]{commenges2013invention}
\bibinfo{author}{Commenges, H.}, \bibinfo{year}{2013}.
\newblock \bibinfo{title}{The invention of daily mobility : Performative
  aspects of the instruments of economics of transportation.}
\newblock \bibinfo{journal}{Theses, Universit{\'e} Paris-Diderot-Paris VII} .
\bibitem[{Goldhirsch et~al.(1987)Goldhirsch, Sulem and
  Orszag}]{goldhirsch1987stability}
\bibinfo{author}{Goldhirsch, I.}, \bibinfo{author}{Sulem, P.L.},
  \bibinfo{author}{Orszag, S.A.}, \bibinfo{year}{1987}.
\newblock \bibinfo{title}{Stability and lyapunov stability of dynamical
  systems: A differential approach and a numerical method}.
\newblock \bibinfo{journal}{Physica D: Nonlinear Phenomena}
  \bibinfo{volume}{27}, \bibinfo{pages}{311--337}.
\bibitem[{Grimm et~al.(2005)Grimm, Revilla, Berger, Jeltsch, Mooij, Railsback,
  Thulke, Weiner, Wiegand and DeAngelis}]{grimm2005pattern}
\bibinfo{author}{Grimm, V.}, \bibinfo{author}{Revilla, E.},
  \bibinfo{author}{Berger, U.}, \bibinfo{author}{Jeltsch, F.},
  \bibinfo{author}{Mooij, W.M.}, \bibinfo{author}{Railsback, S.F.},
  \bibinfo{author}{Thulke, H.H.}, \bibinfo{author}{Weiner, J.},
  \bibinfo{author}{Wiegand, T.}, \bibinfo{author}{DeAngelis, D.L.},
  \bibinfo{year}{2005}.
\newblock \bibinfo{title}{Pattern-oriented modeling of agent-based complex
  systems: lessons from ecology}.
\newblock \bibinfo{journal}{science} \bibinfo{volume}{310},
  \bibinfo{pages}{987--991}.
\bibitem[{Guo and Liu(2011)}]{guo2011bounded}
\bibinfo{author}{Guo, X.}, \bibinfo{author}{Liu, H.X.}, \bibinfo{year}{2011}.
\newblock \bibinfo{title}{Bounded rationality and irreversible network change}.
\newblock \bibinfo{journal}{Transportation Research Part B: Methodological}
  \bibinfo{volume}{45}, \bibinfo{pages}{1606--1618}.
\bibitem[{Han(2003)}]{han2003dynamic}
\bibinfo{author}{Han, S.}, \bibinfo{year}{2003}.
\newblock \bibinfo{title}{Dynamic traffic modelling and dynamic stochastic user
  equilibrium assignment for general road networks}.
\newblock \bibinfo{journal}{Transportation Research Part B: Methodological}
  \bibinfo{volume}{37}, \bibinfo{pages}{225--249}.
\bibitem[{Kryvobokov et~al.(2013)Kryvobokov, Chesneau, Bonnafous, Delons and
  Piron}]{kryvobokov2013comparison}
\bibinfo{author}{Kryvobokov, M.}, \bibinfo{author}{Chesneau, J.B.},
  \bibinfo{author}{Bonnafous, A.}, \bibinfo{author}{Delons, J.},
  \bibinfo{author}{Piron, V.}, \bibinfo{year}{2013}.
\newblock \bibinfo{title}{Comparison of static and dynamic land use-transport
  interaction models: Pirandello and urbansim applications}.
\newblock \bibinfo{journal}{Transportation Research Record: Journal of the
  Transportation Research Board} , \bibinfo{pages}{49--58}.
\bibitem[{Leurent and Boujnah(2014)}]{leurent2014user}
\bibinfo{author}{Leurent, F.}, \bibinfo{author}{Boujnah, H.},
  \bibinfo{year}{2014}.
\newblock \bibinfo{title}{A user equilibrium, traffic assignment model of
  network route and parking lot choice, with search circuits and cruising
  flows}.
\newblock \bibinfo{journal}{Transportation Research Part C: Emerging
  Technologies} \bibinfo{volume}{47}, \bibinfo{pages}{28--46}.
\bibitem[{Mahmassani and Chang(1987)}]{mahmassani1987boundedly}
\bibinfo{author}{Mahmassani, H.S.}, \bibinfo{author}{Chang, G.L.},
  \bibinfo{year}{1987}.
\newblock \bibinfo{title}{On boundedly rational user equilibrium in
  transportation systems}.
\newblock \bibinfo{journal}{Transportation science} \bibinfo{volume}{21},
  \bibinfo{pages}{89--99}.
\bibitem[{Puzis et~al.(2013)Puzis, Altshuler, Elovici, Bekhor, Shiftan and
  Pentland}]{puzis2013augmented}
\bibinfo{author}{Puzis, R.}, \bibinfo{author}{Altshuler, Y.},
  \bibinfo{author}{Elovici, Y.}, \bibinfo{author}{Bekhor, S.},
  \bibinfo{author}{Shiftan, Y.}, \bibinfo{author}{Pentland, A.},
  \bibinfo{year}{2013}.
\newblock \bibinfo{title}{Augmented betweenness centrality for environmentally
  aware traffic monitoring in transportation networks}.
\newblock \bibinfo{journal}{Journal of Intelligent Transportation Systems}
  \bibinfo{volume}{17}, \bibinfo{pages}{91--105}.
\bibitem[{Rasmussen et~al.(2015)Rasmussen, Watling, Prato and
  Nielsen}]{rasmussen2015stochastic}
\bibinfo{author}{Rasmussen, T.K.}, \bibinfo{author}{Watling, D.P.},
  \bibinfo{author}{Prato, C.G.}, \bibinfo{author}{Nielsen, O.A.},
  \bibinfo{year}{2015}.
\newblock \bibinfo{title}{Stochastic user equilibrium with equilibrated choice
  sets: Part ii--solving the restricted sue for the logit family}.
\newblock \bibinfo{journal}{Transportation Research Part B: Methodological}
  \bibinfo{volume}{77}, \bibinfo{pages}{146--165}.
\bibitem[{Sullivan et~al.(2010)Sullivan, Novak, Aultman-Hall and
  Scott}]{sullivan2010identifying}
\bibinfo{author}{Sullivan, J.}, \bibinfo{author}{Novak, D.},
  \bibinfo{author}{Aultman-Hall, L.}, \bibinfo{author}{Scott, D.M.},
  \bibinfo{year}{2010}.
\newblock \bibinfo{title}{Identifying critical road segments and measuring
  system-wide robustness in transportation networks with isolating links: a
  link-based capacity-reduction approach}.
\newblock \bibinfo{journal}{Transportation Research Part A: Policy and
  Practice} \bibinfo{volume}{44}, \bibinfo{pages}{323--336}.
\bibitem[{Tordeux and Lassarre(2016)}]{tordeux2016jam}
\bibinfo{author}{Tordeux, A.}, \bibinfo{author}{Lassarre, S.},
  \bibinfo{year}{2016}.
\newblock \bibinfo{title}{Jam avoidance with autonomous systems}.
\newblock \bibinfo{journal}{arXiv preprint arXiv:1601.07713} .
\bibitem[{Tsai(2005)}]{tsai2005quantifying}
\bibinfo{author}{Tsai, Y.H.}, \bibinfo{year}{2005}.
\newblock \bibinfo{title}{Quantifying urban form: compactness versus' sprawl'}.
\newblock \bibinfo{journal}{Urban studies} \bibinfo{volume}{42},
  \bibinfo{pages}{141--161}.
\bibitem[{Wardrop(1952)}]{wardrop1952road}
\bibinfo{author}{Wardrop, J.G.}, \bibinfo{year}{1952}.
\newblock \bibinfo{title}{Some theoretical aspects of road traffic research.}
\newblock \bibinfo{journal}{Proceedings of the institution of civil engineers}
  \bibinfo{volume}{1}, \bibinfo{pages}{325--362}.
\bibitem[{Zhang et~al.(2013)Zhang, Mahmassani and Lu}]{zhang2013dynamic}
\bibinfo{author}{Zhang, K.}, \bibinfo{author}{Mahmassani, H.S.},
  \bibinfo{author}{Lu, C.C.}, \bibinfo{year}{2013}.
\newblock \bibinfo{title}{Dynamic pricing, heterogeneous users and perception
  error: Probit-based bi-criterion dynamic stochastic user equilibrium
  assignment}.
\newblock \bibinfo{journal}{Transportation Research Part C: Emerging
  Technologies} \bibinfo{volume}{27}, \bibinfo{pages}{189--204}.
\bibitem[{Zhu and Levinson(2010)}]{zhu2010people}
\bibinfo{author}{Zhu, S.}, \bibinfo{author}{Levinson, D.},
  \bibinfo{year}{2010}.
\newblock \bibinfo{title}{Do people use the shortest path? an empirical test of
  wardrop's first principle}, in: \bibinfo{booktitle}{91th annual meeting of
  the Transportation Research Board, Washington},
  \bibinfo{organization}{Citeseer}.

\end{thebibliography}
\bibliographystyle{elsarticle-harv}

\end{document}